%% file: batterio.tex
\documentclass[a4paper,11pt,fleqn]{article}
\usepackage{latexsym}
\usepackage{amsmath}
\usepackage{amssymb}
\usepackage[dvips]{graphicx}
\author{  \quad \\ Luigi  Sertorio $^{\dag}$  \and   \quad \\ Giovanna  Tinetti 
$^{\ddag}$ }
\title{\textbf{{\LARGE Available Energy for Life on a Planet, with or without
Stellar Radiation.}}}
\date{  {\footnotesize \emph{Dipartimento di Fisica Teorica -
Universit\`a di Torino. \\ Istituto Nazionale di Fisica Nucleare -
Sezione di Torino}} \\ \quad \\ {\footnotesize DFTT 33/00} \\ \quad \\
{\footnotesize $^{\dag}$ sertorio@to.infn.it \qquad \qquad  
 $^{\ddag}$  tinetti@to.infn.it } }
\begin{document}
\maketitle
\renewcommand{\theequation}{\thesection.\arabic{equation}}
\renewcommand{\thefigure}{\thesection.\arabic{figure}}
\newcommand{\ud}{\mathrm{d}}
\newcommand{\de}{\partial}
\input{batab}
\include{batt1}

\include{batt2}

\include{batt3}
\include{batt4}

\include{batt5}

\include{batt6}

\include{batt7}

\include{batt8}

\include{biblio}
\end{document}

%% file: batab.tex
\section*{Abstract}
\begin{tabular}{|p{14. cm}|}
\hline
{\small
The quest for life in the Universe is often affected by the free use of extrapolations of our phenomenological geocentric 
knowledge.
We point out that the existence of a living organism, and a population of organisms, requires the existence of available energy or,
more precisely, available power per unit volume (\S 1). This is not a geocentric concept, but a principle that belongs to the foundations of thermodynamics. A quest about availability in the Universe is justified. We discuss the case in which power comes from mining (\S 2),
and from thermal disequilibrium (\S 3). Thermal disequilibrium may show up in two ways: on planets without a star (\S 4),
and on planets where the surface thermal disequilibrium is dominated by the incoming photon flux from the nearest star (\S 6).
In the first case we study the availability by simulating the structure of the planet with a simple model
that contains the general features of the problem. For the first case we show that the availability is in general very small (\S 5).
In the second case we show that the availability is in general large; the order of magnitude depends first of all on the star's temperature and the planet's orbit, but is also controlled by the greenhouse gases present on the planet. 
\par }
    \\
\hline
\end{tabular}   \newpage

%% file: batt1.tex
\section{Introduction}
\setcounter{figure}{0}
\setcounter{equation}{0}
We consider the question of whether there is life elsewhere in the Universe, and not only on Earth. This question is complicated by the fact that we do not have a definition of life as we do for matter, the object of physics: the laws of physics refer to something that is the same everywhere and at all times.
\\
For life we need an ad hoc hypothesis. For instance,  we can assume that life is a phenomenon unique to the  Earth or
that life has a uniform chance to exist in the Universe, provided certain conditions are satisfied. We take the second hypothesis \cite{ref:hbd}.
With regards to  the conditions, we do have several descriptions of living organisms on Earth, hence  conditions that are nevertheless local and possibly restrictive. In fact a certain kind of life, somewhere, could be different in appearance and performance, adapted to 
non-terrestrial environments.
\\
In the absence of a general definition that would resolve these questions, we limit ourselves to discuss the most general requirement: 
available energy.
To this purpose, we adopt this weak definition of life: a living organism occupies a finite domain, has structure, performs according to an unknown purpose, and reproduces itself; in order to satisfy these requirements it uses a certain amount of power density  
($\frac{\text{W}}{\text{m}^{3}}$).  
That a supply of power is necessary, is obvious, upon considering what a living organism is not.\\
 A reversible event does not  require power in order to happen; it only requires an initial condition, the Hamiltonian does the rest.
\\
A dissipative structure requires power; examples are the Hadley cells \cite{ref:peo} in a planetary atmosphere, which contain 
ongoing cyclic motions plus chemical reactions, and, in the laboratory, the B\'enard cells \cite{ref:cha}, the 
Belousov Zhabotinsky cycle \cite{ref:bru}, etc.
 These physical structures have a property: by changing the energy flow across a certain boundary, the structure may appear or 
disappear: there is a bifurcation in a dynamical dissipative system. \\
A living organism is not like this, because it is a control system: its structure is related to the operation of controls, that
 are actuated according to an unknown purpose. The purpose includes the preservation of the structure itself, plus  reproduction, plus a certain 
performance.
These concepts exist also in physics and engineering, in  that branch called control theory. For example a spaceship contains several auxiliary jet engines, that are operated according to a certain strategy of flight. The operation of such engines has the effect of changing, at appropriate instants of time $t_{i}$,
the initial condition, or phase space configuration at  times $t_{i}$,
of the Newtonian equations of motion of the spaceship; this behaviour is finalistic rather than deterministic, even if at all times the spaceship obeys
the deterministic laws of gravity. Similarly the living organism performs according to a finalistic strategy although at all times obeys the laws of physics and chemistry. Both the spaceship and the living organism require a  power supply.
\\
In conclusion we make only the  hypothesis that the energy requirements for life is
the same everywhere in the Universe.
\\
Closely related to this hypothesis is another question that concerns the material flow processed by the living organism, 
once the power is supplied. In fact a living organism receives energy and rejects energy, receives chemical elements and rejects chemical elements.
The chemical elements are extracted and rejected from and into the same physical ambient.
Such a process can be open, or closed through an ambient cycle connected to the chemical processes occurring in the inside
of the living organism. This cycle exists on the planet Earth. We do not know much about this cycle, therefore it is not clear
how its existence can be incorporated as an explicit requisite for life everywhere. \\
We discuss briefly the absence of the cycle in \S 2, dedicated to mining, and we limit ourselves in the other
 sections to the evaluation of the available energy. \\ 

Last remark. 
In the absence of a general definition of life, we have adopted in this paper a weak definition that a living organism is finite in space, has structure, performs, and reproduces itself. Hence life has to be \emph{chemical life}, and that is hardly conceivable not in connection to a planet. \\
An even weaker definition of life is no definition; in this case the search for available energy can be pushed to the extreme condition that can be guessed in the asymptotic behaviour of the cosmic expansion  \cite{ref:dys}, \cite{ref:kra}. The available energy, and relative information, come from strategies of mining, all in the vicinity of the cosmic background temperature.
We call this \emph{abstract life}. \\
It is not clear if an intelligence-preserving transition between chemical life and abstract life is conceivable in principle.

%% file: batt2.tex
\section{Energy by mining}
\setcounter{figure}{0}
\setcounter{equation}{0}
The unit organism occupying a volume $V$ needs a certain power $\varphi$. A population of $n (t)$
organisms requires a power 
\begin{equation*}
\dot{q} (t) = n (t) \, \varphi; \qquad \qquad \qquad ( [\dot{q}] = \text{W}) 
\end{equation*}
The population number $n (t)$ is a variable because typical of life is  reproduction. The time dependence $n (t)$
is determined by the available resources and by the collective strategies of evolution of the living population.
Mining for energy is a well known concept to humankind for about two centuries; that is since the time of mining energy from coal and petroleum reservoirs. The population of living organisms functioning according to some unknown strategy, extracts from its neighborhood the incoming chemical elements $a, \, b, \, c \ldots $ feeds its own complex state
$C$ with both mass and energy and ejects the outgoing chemical elements $\alpha, \, \beta, \, \gamma, \ldots $
to another neighborhood. 
\begin{center}
\includegraphics[width=8 cm]{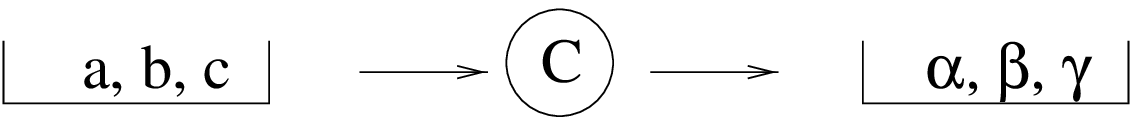}
\end{center} 
The energy balance is described by the following figure:
\begin{center}
\includegraphics[width=9 cm]{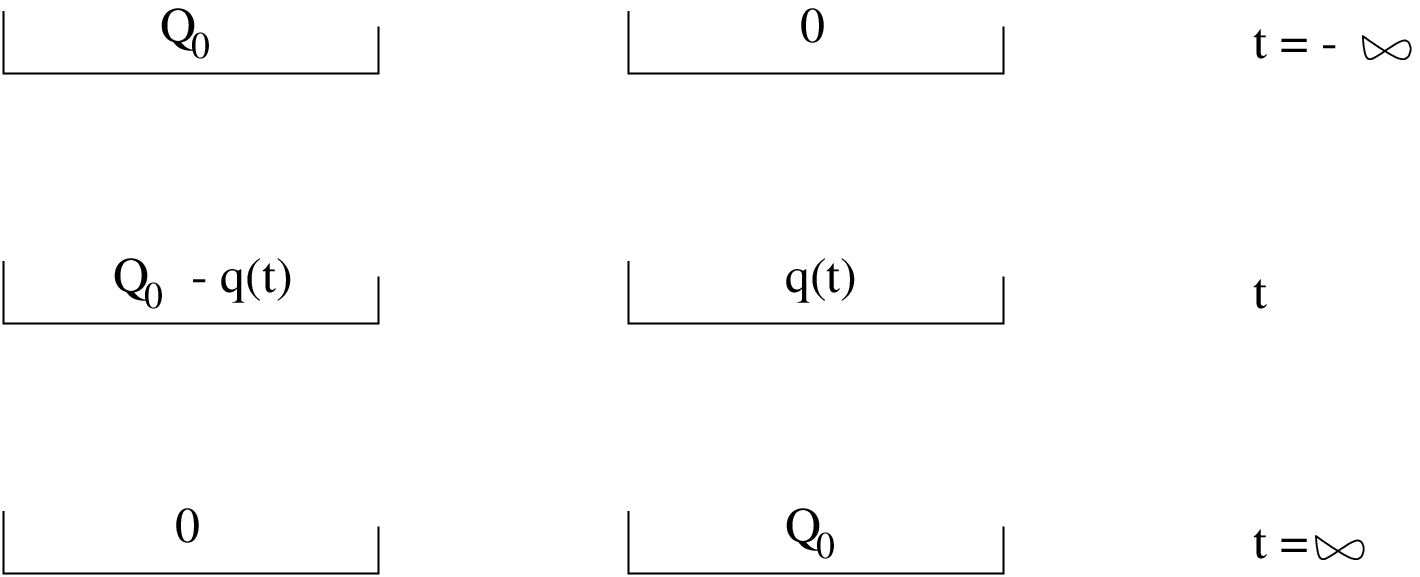}   
\end{center}
$Q_{0}$ and $q (t)$ are expressed in Joule. \\
Next we ask which time scales are pertinent to the permanence of the population $n (t)$.
We wish not to commit ourselves to wild hypothesis on the methodology of proliferation of the unknown form of life, the complex state
$C$, and its population $n (t)$. The minimum hypothesis is the hypothesis of logistic growth \cite{ref:log}. In this case $q (t)$, obeys the logistic equation
\begin{equation} \label{eq:ii1}
\dot{q} = a \, q \, \frac{Q_{0} - q}{Q_{0}}; \qquad \qquad \qquad [q] = \text{J}
\end{equation}
which has the solution
\begin{equation} \label{eq:ii2}
q =  \frac{Q_{0}}{1 + e^{- a \, t} \, \left ( \frac{Q_{0}}{q (0)} - 1 \right )}
\end{equation}
The graph of (\ref{eq:ii2}) is of the kind
\begin{center}
\includegraphics[width=8 cm]{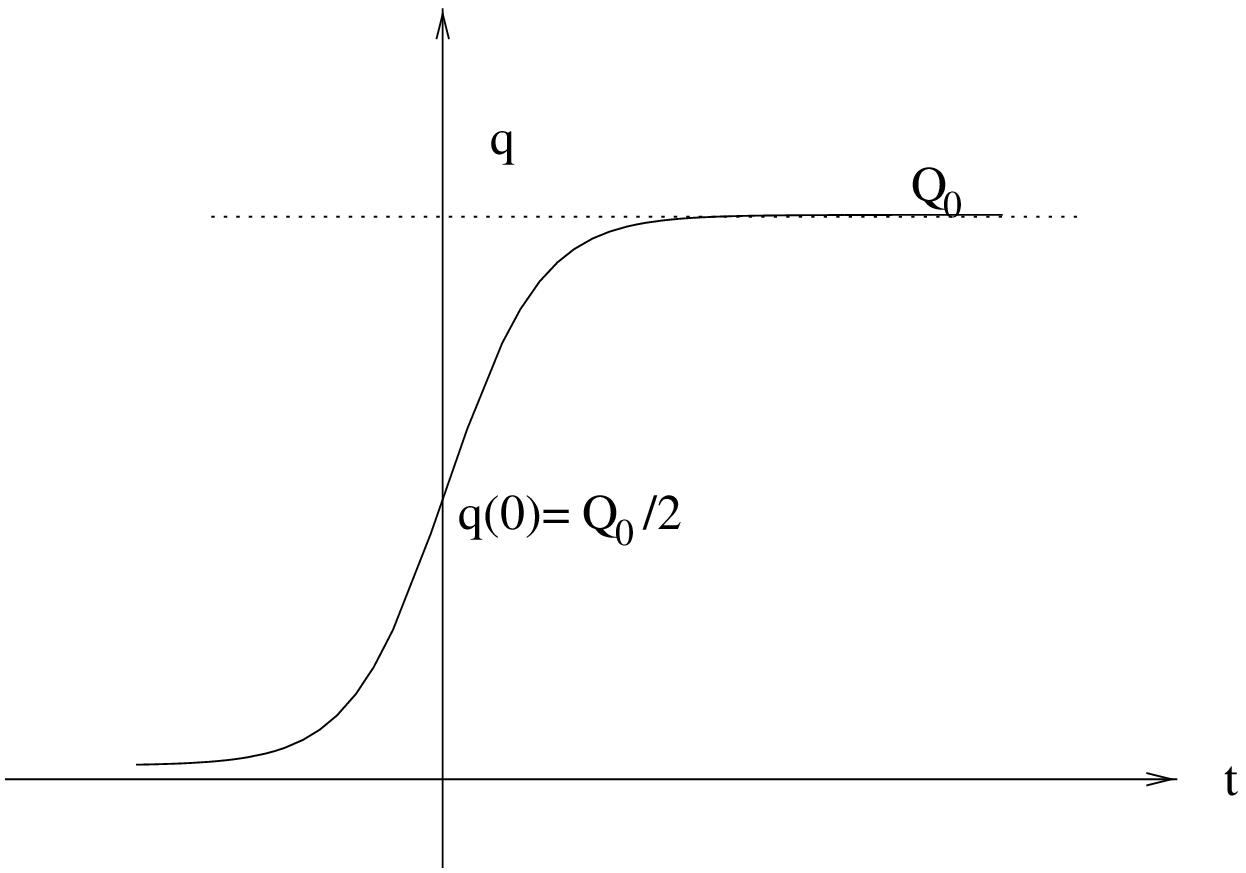}   
\end{center}
The collective growth parameter $a$ is the inverse of the time scale $\tau$,
\begin{equation} \label{eq:ii3}
a = \frac{1}{\tau}.
\end{equation}
The power used by the population is
\begin{equation} \label{eq:ag}
\dot{q} = n (t) \, \varphi; \qquad \qquad \qquad [\varphi] = \text{W}
\end{equation}
and is obtained by derivation with respect to the time of (\ref{eq:ii2})
\begin{equation} \label{eq:ii4}
\dot{q} =  \frac{a \, Q_{0} \, e^{- a \, t} \, \left ( \frac{Q_{0}}{q (0)} - 1 \right )}{\left
 [ 1 + e^{- a \, t} \, \left ( \frac{Q_{0}}{q (0)} - 1 \right ) \right ]^{2}}
\end{equation}
The behaviour of (\ref{eq:ii4}) is of the kind
\begin{center}
\includegraphics[width=8 cm]{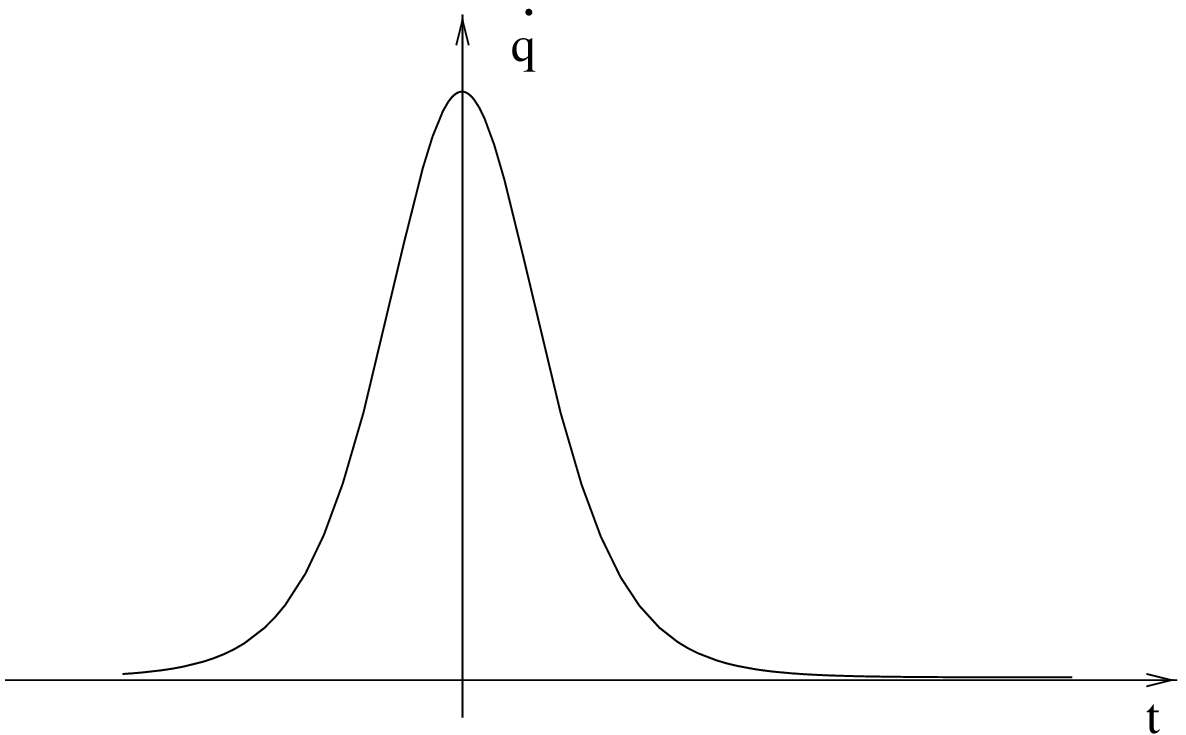}   
\end{center}
Notice that
\begin{equation} \label{eq:ii5}
\dot{q}^{max} = \frac{Q_{0}}{4 \, \tau}
\end{equation}
Therefore there is a trade-off between mean life $\tau$ and maximum population $n^{max}$. From (\ref{eq:ag}) and (\ref{eq:ii5}): 
\begin{equation} \label{eq:ii6}
n^{max} = \frac{Q_{0}}{4 \, \tau \, \varphi}
\end{equation}
A violation of the rule (\ref{eq:ii6}) can very well be imagined; for some unknown reason the population growth may not obey the logistic equation (\ref{eq:ii1}), which implies inexorably the asymptotic exponential disappearance of life. 
A particular population of living organisms may escape the logistic rule and choose survival rather than growth.
This would be a very interesting life form to discover. \\
As a conclusion, we emphasize that the time of survival depends on the strategy of the living community itself, is not a property of the ambient in which the hypothetical life form develops.

%% file: batt3.tex
\section{Access to thermal power}
\setcounter{equation}{0}
\setcounter{figure}{0}
Next we consider an ambient  which provides  a local flow of power. 
A local flow of power is embedded in a standing, non uniform temperature and 
pressure distribution.  For simplicity, we will consider in the foregoing sections only a temperature distribution. 
Two cases can be considered:
a non uniform distribution of temperature can be a transient, from an initial state to a final state, or can be a steady, non equilibrium state. \\
A natural realization of the first case is a planet without star during the thermodynamical process  of cooling off. In fact we may consider the situation whereby after several events that brought the planet to its formation and maturity, the structure is relatively fixed and the thermal flow develops with time, as in an initial
condition problem, giving rise to a long transient of cooling. \\
A natural realization of the second case is a planet with a star; it receives
a flow of radiation from the star, and if this flow can be considered constant
(for this the orbit must be circular) the temperature distribution on the planet's surface, which is never uniform, is nevertheless globally stationary in the sense that the total incoming radiation is equal to the total outgoing radiation, and both are constant. If the orbit is far from circular, the total incoming flow is no longer constant and the temperature distribution can be very complicated. In all cases, any non equilibrium distribution of temperature (and pressure) contains in itself a certain amount of ``available energy''. This concept has been formulated by Gibbs in equilibrium thermodynamics
\cite{ref:gib}, and extended by Landau to the formalism of non equilibrium thermodynamics  \cite{ref:lan}.
Here follows a brief review of  the concept of Gibbs availability.
\\
To any situation of thermodynamical disequilibrium there corresponds a well defined amount of energy. This is the minimum energy required, in principle, to produce the disequilibrium, the uniform equilibrium being the natural state; or, conversely, this is the maximum energy which, in principle, 
could be extracted exploiting the given disequilibrium.
The words ``maximum/minimum in principle'' mean that either operation is done reversibly. Let us see how. 
\\
In equilibrium thermodynamics the concept of Gibbs availability comes by considering a closed system in a volume $V_{0}$
with equilibrium values of pressure and temperature respectively 
$P_{0}$ and $T_{0}$; and a portion of volume $V < V_{0}$ which 
is out of equilibrium with $P > P_{0}$,  $T > T_{0}$.
This disequilibrium situation has been created reversibly by feeding this amount of energy:
\begin{equation} \label{eq:iii1}
\Delta W = \Delta Q \, \frac{T - T_{0}}{T} + \Delta V \, (P - P_{0})
\end{equation}
\begin{tabular}{ll}
$\Delta Q$ & is incoming, as seen from the subsystem $V$, \\
& is outgoing, as seen from the system $V_{0}$. \\
\end{tabular} \\
\begin{tabular}{ll}
$\Delta V$ & is volume reduction, as seen from the subsystem $V$, \\
& is volume expansion, as seen from the system $V_{0}$. \\
\end{tabular} \\ \quad \\
Notice that $\frac{T - T_{0}}{T} = \eta$ is the Carnot efficiency. \\
In the natural return to equilibrium the quantity $\Delta W$ is entirely 
dissipated, but if the disequilibrium situation is undone reversibly, 
$\Delta W$ could be extracted. In this sense the minimum work $W$ becomes 
the maximum availability. \\
In non equilibrium thermodynamics we do not consider two pieces $V$ and $V_{0}$, but a distribution $T (\vec{r}, t), \, P (\vec{r}, t)$. The generalization of (\ref{eq:iii1}) is intuitive. Rather than a quantity of energy $\Delta W$ we consider a power density
\begin{equation*}
w = \frac{\ud W}{\ud V \, \ud t}
\end{equation*}
The natural extension of (\ref{eq:iii1}) is
\begin{equation} \label{eq:iii2}
w = w^{carnot} + w^{fluid} =  \vec{q} \cdot \vec{\eta} + \nabla P \cdot \vec{v}; \qquad \qquad \qquad [w] = \frac{\text{W}}{\text{m}^{3}}
\end{equation}
In (\ref{eq:iii2}) 
\begin{equation*}
 \begin{array}{ll}
\nabla & \text{is the gradient operator} \\
\vec{v} & \text{is the fluid velocity} \\
\vec{\eta} = \frac{\nabla T}{T} & \\
\vec{q} = \frac{\ud Q}{\ud S \, \ud t} & \text{is the heat flow}, \qquad \qquad [\vec{q}] = \frac{\text{W}}{\text{m}^{2}}\\
\end{array} \end{equation*} 
It is easy to convince oneself that (\ref{eq:iii2}) is correct. Consider the volume $\ud V = \ud S \, \ud \ell$
\begin{figure}[ptbh]
\begin{center}
\includegraphics[width=3.5 cm]{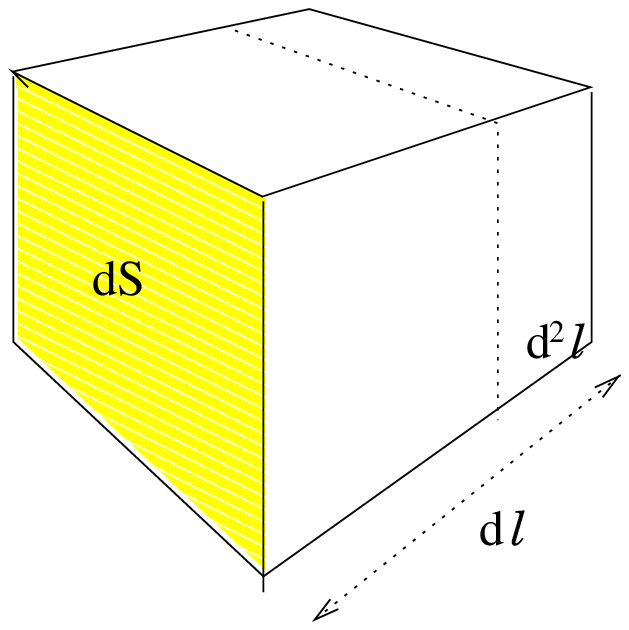}   
\caption{} \label{fig:ag}
\end{center} 
\end{figure}
During the time $\ud t$ the increment $\ud Q$ pertinent to $\ud V$ is $\frac{\ud Q}{\ud t}$ and the Carnot efficiency is $\eta = \frac{\ud T}{T}$, where $\ud T$ is the temperature change along $\ud \ell$. Here $\eta$ is the usual dimensionless
efficiency. \\
We can write
\begin{equation} \label{eq:iii3}
w^{carnot} = \frac{\ud Q}{\ud t} \, \frac{\ud T}{\ud S \, \ud \ell} \, \frac{1}{T} = \vec{q} \cdot \vec{\eta}
\end{equation}
In (\ref{eq:iii3})
\begin{equation*} 
 \vec{\eta} = \frac{\nabla T}{T}; \qquad \qquad \qquad [\vec{\eta}] = \text{m}^{-1}
\end{equation*}
and $\vec{q} = \frac{\ud Q}{\ud S \, \ud t}$ is the common definition of heat flow.
In terms of the distribution $T (\vec{r}, t)$, equation (\ref{eq:iii3}) becomes
\begin{equation} \label{eq:iii4}
w^{carnot} = k \, \frac{(\nabla T )^{2}}{T} 
\end{equation}
We have simply used the definition, valid for heat conduction,
\begin{equation*}
\vec{q} = - k \, \nabla T
\end{equation*}
where $k$ is the Fourier constant. \\
Remember that if we neglect convective phenomena, the non equilibrium dynamics is governed by the Fourier heat equation
\begin{equation*}
c_{v} \, \rho \, \frac{\de T}{\de t} = - \text{div} \, \vec{q} =  k \, \nabla \cdot \nabla T = k \, \Delta T
\end{equation*}
$k$ is the Fourier constant mentioned above and $e = c_{v} \, \rho \, T$ is the energy equation of state. \\
For the pressure part, using the same $\ud V$, we have, during the time $\ud t$
\begin{equation*}
\ud P \, \ud S \, \ud^{2} \ell = \ud W; \qquad \qquad \qquad [\ud W] = \text{J}
\end{equation*}
$\ud^{2} \ell$ is the contraction of $\ud \ell$ (see fig. \ref{fig:ag}), or, equivalently,
$\ud S \, \ud^{2} \ell = \ud^{2} V$
the contraction of $\ud V$. There follows 
\begin{equation} \label{eq:iii5}
w^{fluid} = \frac{\ud W}{\ud S \, \ud \ell \, \ud t} = \nabla P \cdot \vec{v};
\qquad \qquad \qquad [\nabla P \cdot \vec{v}] =  \frac{\text{W}}{\text{m}^{3}}
\end{equation}
and (\ref{eq:iii2}) is justified. \\
In the next sections we will limit ourselves on the consideration of $w^{carnot}$, equation (\ref{eq:iii4}).

%% file: batt4.tex
\section{Planet without star. The transient}
\setcounter{equation}{0}
\setcounter{figure}{0}
A planet without incoming radiation can belong to a star on a very distant orbit, or to a double star on an outer orbit, 
or can be a planet traveling in some pan-galactic orbit.
\\
For such a planet there is the possibility of containing a non equilibrium global transient. In fact, following the idea that planets are formed by accretion \cite{ref:alf}, we may consider a stage of maturity in which the structure is stabilized and the initial temperature is generally a distribution which decreases from the center to the periphery. This is a universal feature that has two reasons to occur. \\
The pressure must be higher in the deep interior with respect to the surface, due to the force of gravity; on the other hand,
the two response functions  $\left ( \frac{\de U}{\de T}  \right )_{V}$ and $\left ( \frac{\de P}{\de T}  \right )_{V}$ 
must be positive, because of the second law.
So energy and pressure increase with $T$, therefore the combination of gravity and the second law induces a hot interior. 
\\
The second reason is that the collisions during the process of accretion develop heat, and these violent events are followed by a 
spontaneous creation of layers.
In fact the various materials  present in the planet  have in general different heat conduction parameters; the good conductors keep transferring heat,  the bad conductors remain on the surface $(T \sim 0)$, where they cool off but do not transfer
much heat and therefore
 constitute the insulating crust.
This stratification can be quite complicated, although it is a general feature.\\
 Here we consider the simple model of a core perfect conductor, and an insulating crust.
The external surface cools off with the Stefan-Boltzmann law
\begin{equation*}
q = \sigma \, T^{4}; \qquad \qquad \qquad [q] = \frac{\text{W}}{\text{m}^{2}}
\end{equation*}
$q$ is the heat flow and $\sigma \simeq 5.67 \cdot 10^{-8} \, \frac{\text{W}}{\text{m}^{2} \, \text{K}^{4}} $ is the Stefan-Boltzmann constant.
Actually the heat emitted by radiation depends on the nature of the surface, which is not necessarily a black solid.
If the surface is solid but not black, we may write
\begin{equation} \label{eq:iv0}
q = \epsilon \, \sigma \, T^{4}
\end{equation}
where $\epsilon$ is very close to 1. 
\\
If the surface is gaseous, its molecules may undergo events of inelastic scattering with the photons originated by the solid part 
below.
So the gas layer acts as an insulating greenhouse atmosphere. 
Nevertheless, in the absence of a star,  the surface will eventually become  cold and 
the gas phase will not be sustained and we return to (\ref{eq:iv0}) with $\epsilon \sim 1$ valid for solids. \\
A general analytic treatment, even assuming spherical symmetry of the temperature field, is very complicated. 
In particular, it requires a guess on the initial condition $T (r, t=0)$; and this has a large  uncertainty; secondly it requires a guess on the external surface behaviour of $\nabla T$, and this is again open to uncertainty. \\
Our choice is to make the ``minimum guess'' compatible with the coexistence of two problems: what gives the time scale of the cooling process and which is the order of magnitude of the Gibbs availability? We anticipate that we will find a relationship between availability and lifetime of the transient at the end of \S 5. \\
The model planet we are considering is shown in this figure \\
\begin{center}
\includegraphics[width=4.4 cm]{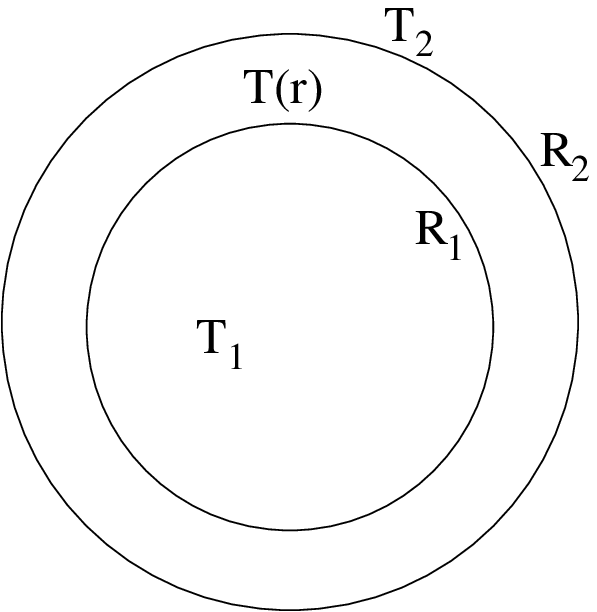}   
\end{center}
\begin{equation*}
\begin{array}{ll}
 T = T_{1} (t) &   0 \le r \le R_{1} \\
T = T (r) & R_{1} < r < R_{2} \\
T = T_{2} (t) & r = R_{2}  \\
\end{array}
\end{equation*}
The core contains the thermal energy reservoir at uniform temperature $T_{1} (t)$. The insulating layer produces $\nabla T (r)$
and consequently gives implicitly $T_{2} (t)$.
The radial flow across the shell is 
\begin{equation} \label{eq:iv1}
k \, r^{2} \, \frac{\de T}{\de r} = \frac{\ud \phi}{\ud \Omega} = \text{const.}  \qquad \qquad \qquad [\phi] = \text{W}
\end{equation}
$\ud \Omega$ is the solid angle
\begin{equation*} 
\ud \Omega = \frac{\ud S (r)}{r^{2}}
\end{equation*}
We evaluate $\frac{\ud \phi}{\ud \Omega}$ at $r = R_{2}$
\begin{equation*} 
 \frac{\ud \phi}{\ud \Omega} = R_{2}^{2} \,  \frac{\ud \phi}{\ud S (R_{2})} = \sigma \, T_{2}^{4} \, R_{2}^{2}
\end{equation*}
so that (\ref{eq:iv1}) becomes
\begin{equation} \label{eq:iv2}
k \, r^{2} \, \frac{\de T}{\de r} = \sigma \, T_{2}^{4} \, R_{2}^{2}
\end{equation}
The function $T (r)$ is obtained integrating (\ref{eq:iv2}).
We get
\begin{equation} \label{eq:iv3}
T  (r) = T_{1} - \frac{1}{k} \, \sigma \, T_{2}^{4} \, R_{2}^{2} \, \, \frac{r - R_{1}}{r \, R_{1}}
\end{equation}
Notice that $T_{1}$ and $T_{2}$ are at this point unknown.
They are related by the following algebraic equation, arrived at by  putting $r = R_{2}$ in (\ref{eq:iv3})
\begin{equation} \label{eq:iv4}
T_{2} = T_{1} - a \, T_{2}^{4}
\end{equation}
where $a$ is given by
\begin{equation} \label{eq:iv5}
a = \frac{\sigma}{k}  \, \frac{R_{2}}{R_{1}} \,  \Delta R \qquad \qquad \qquad [a] = \text{K}^{-3}
\end{equation}
and $\Delta R = R_{2} - R_{1}$.
The time dependence $T_{1} (t)$ and $T_{2} (t)$ is given by the energy balance equation:
\begin{equation} \label{eq:iv6}
\frac{4}{3} \, \pi \, R_{1}^{3} \, c_{v} \, \rho \,  \dot{T}_{1} (t) =
- 4 \, \pi \, R_{2}^{2} \, \sigma \, T_{2}^{4} (t).
\end{equation}
This equation simply relates the energy loss from the volume $V_{1} = \frac{4}{3} \, \pi \, R_{1}^{3} $ to the energy flow across the surface $4 \, \pi \, R_{2}^{2}$.
Equation (\ref{eq:iv4}) gives $T_{1}$ as a function of $T_{2}$
\begin{equation} \label{eq:iv7}
T_{1} = T_{2} + a \, T_{2}^{4}; 
\end{equation}
putting (\ref{eq:iv7}) into (\ref{eq:iv6}) we get
\begin{equation} \label{eq:iv8}
\frac{1}{3} \, c_{v} \, \rho  \, R_{1}^{3} \, \dot{T}_{2}  \, \left ( 1 + 4 \, a \, T_{2}^{3}  \right  ) =
-  R_{2}^{2} \, \sigma \, T_{2}^{4}, 
\end{equation}
which is a first order differential equation in $T_{2} (t)$ with initial condition
\begin{equation*} 
 T_{2} (0) = T_{20}.
\end{equation*}
We may like to take for initial condition the temperature of the core; from (\ref{eq:iv7}) we may set
\begin{equation*} 
 T_{1} (0) = T_{10}
\end{equation*}
but it is harder to get $T_{2}$ from $T_{1}$. \\
Giving our attention to $T_{1}$ means emphasizing the accretion heating and the total mass of the planet. 
Giving our attention to $T_{2}$ means emphasizing the properties of the crust, namely its thickness $\Delta R$ and its conductivity. \\
Equation (\ref{eq:iv8}) is solved by  separation of the variables, $T_{2}$ and $t$. The solution is 
\begin{equation} \label{eq:iv9}
 t = - 4 \,   a \, b \, \ln \frac{T_{2}}{T_{20}} + \frac{b}{3}  \,  \left ( \frac{1}{T_{2}^{3}} - \frac{1}{T_{20}^{3}} 
\right ) 
\end{equation}
where $a$ is given by (\ref{eq:iv5})
and $b$ is given by
\begin{equation} \label{eq:iv10}
b = \frac{c_{v} \, \rho  \, R_{1}^{3}}{ 3 \, \sigma \, R_{2}^{2}}; \qquad \qquad [b] = \text{K}^{3} \cdot \text{s}
\end{equation}
Equation (\ref{eq:iv9}) implicitly  gives  $T_{2} (t; T_{20})$, or, in brief, $T_{2} (t)$; and using (\ref{eq:iv7})
we have $T_{1} (t; T_{20})$, or $T_{1} (t)$. With this knowledge, equation (\ref{eq:iv3}) is now completely determined and we 
can write
\begin{equation} \label{eq:iv11}
T  (r, t) = T_{1} (t; T_{20}) - \frac{1}{k} \, \sigma \, T_{2}^{4} (t; T_{20})
\, R_{2}^{2} \, \, \frac{r - R_{1}}{r \, R_{1}}
\end{equation}
Consider the mean life of $T_{2} (t)$ coming from (\ref{eq:iv9}), namely the time $\tau$ such that
\begin{equation} \label{eq:iv12}
T_{2} (\tau) = e^{-1} \, T_{20}
\end{equation}
We get
\begin{equation} \label{eq:iv13}
\tau = 4 \, a \, b  + \frac{1}{3} \,(e^{3} - 1) \, b \, \frac{1}{T_{20}^{3}}   = \tau_{1} + \tau_{2}
\end{equation}
The first addend $\tau_{1}$ contains only the planet's parameters, while $\tau_{2}$ contains also the initial condition.
This suggests that the differential equation (\ref{eq:iv8}) has two regimes. In fact, consider the domains of $T_{2}$ such that
\begin{equation} \label{eq:iv14}
\begin{array}{llr}
T_{2} > & \displaystyle{ \left ( \frac{1}{4 \, a}  \right )^{\frac{1}{3}}   } &  \qquad \qquad \qquad
\qquad \qquad \qquad \qquad  a) \\
& & \\
T_{2} < & \displaystyle{ \left ( \frac{1}{4 \, a}  \right )^{\frac{1}{3}}   } &  \qquad \qquad \qquad b) \\
\end{array}
\end{equation}
In the range (\ref{eq:iv14} a), equation (\ref{eq:iv8}) is approximated by
\begin{equation} \label{eq:iv15}
\frac{4}{3} \,a \,  c_{v} \, \rho  \, R_{1}^{3} \, \dot{T}_{2}   =
-  R_{2}^{2} \, \sigma \, T_{2} 
\end{equation}
The solution of (\ref{eq:iv15}) is 
\begin{equation} \label{eq:iv16}
T_{2} (t) = T_{2} (0) \, e^{- \frac{t}{\tau_{1}}}  
\end{equation}
with
\begin{equation} \label{eq:iv17}
\tau_{1} = 4 \, a \, b  = \frac{4}{3} \,   \frac{c_{v} \, \rho}{k}  \, \Delta R \, \frac{R_{1}^{2}}{R_{2}}
\end{equation}
The value
\begin{equation} \label{eq:iv18}
T_{2}^{*} =  \left ( \frac{1}{4 \, a}  \right )^{\frac{1}{3}}  =
 \left ( \frac{k \, R_{1}}{4 \, \sigma \, R_{2} \,  \Delta R}  \right )^{\frac{1}{3}}
\end{equation}
is relevant also for equation (\ref{eq:iv7}): in fact in the range (\ref{eq:iv14} b) the relationship between $T_{1}$ and $T_{2}$ is linear, while
in the range (\ref{eq:iv14} a) the relationship is dominated by the behaviour 
\begin{equation} \label{eq:iv19}
T_{1} \sim a \, T_{2}^{4}; 
\end{equation}
In the range (\ref{eq:iv14} b) equation (\ref{eq:iv8}) is approximated by
\begin{equation} \label{eq:iv20}
\frac{1}{3} \, c_{v} \, \rho  \, R_{1}^{3} \, \dot{T}_{2}   =
-  R_{2}^{2} \, \sigma \, T_{2}^{4} 
\end{equation}
with solution
\begin{equation} \label{eq:iv21}
T_{2} (t) = \left ( \frac{1}{T_{20}^{3}} + \frac{3}{b} \, t \right )^{- \frac{1}{3}}  
\end{equation}
and
\begin{equation} \label{eq:iv22}
\tau = \tau_{2} = \frac{1}{9} \, (e^{3} - 1) \, \frac{c_{v} \, \rho}{\sigma}  \, \frac{R_{1}^{3}}{R_{2}^{2}} \, \frac{1}{T_{20}^{3}}
\end{equation}
Remember that in this model the planet is never at uniform temperature for $R_{1} \le r \le R_{2}$, but  the assumed 
initial distribution is $T_{1} \gg T_{2}$, so in general it is the exponential solution (\ref{eq:iv16}) that interests us. \\
We show in fig. \ref{fig:iv1} the function $T_{1} = T_{1} (T_{2})$ or $T_{2} = T_{2} (T_{1})$ for a given choice of the parameters 
$k$ and $ \Delta R$; $\sigma$ is a fundamental constant.
The region $T_{2} < T_{2}^{*}$ is magnified. \\
\begin{figure}[ptbh]
\begin{center}
\includegraphics[width=5.2 cm,bbllx=4bp,bblly=4bp,bburx=202bp,bbury=287bp,
clip=]{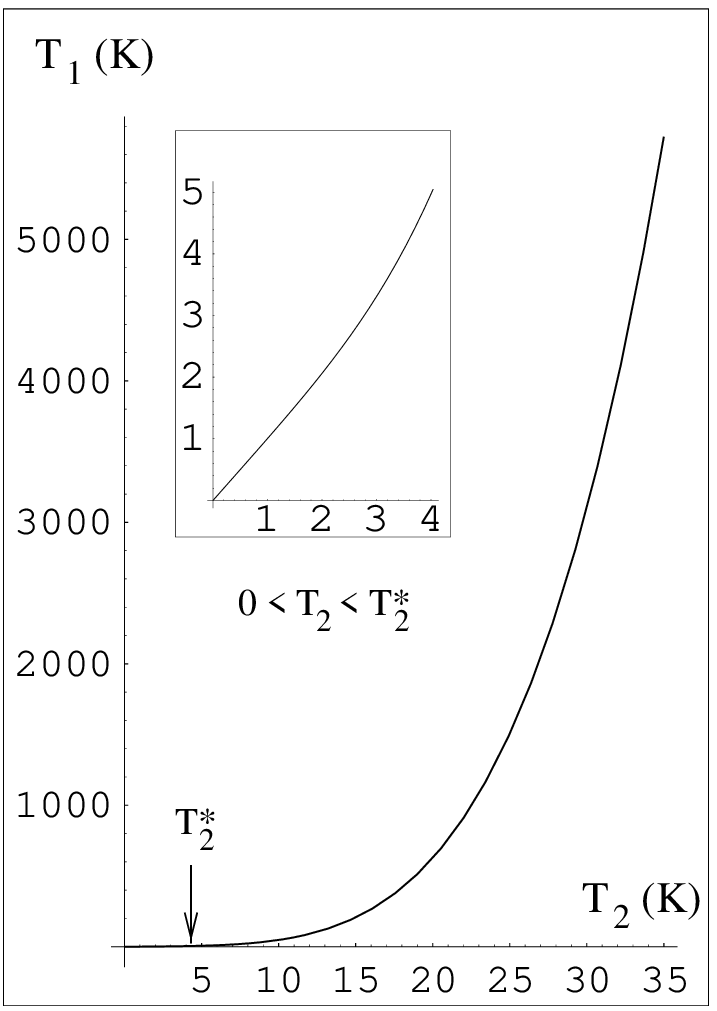} \quad \quad
\includegraphics[width=7.5cm,bbllx=4bp,bblly=6bp,bburx=286bp,bbury=185bp,
clip=]{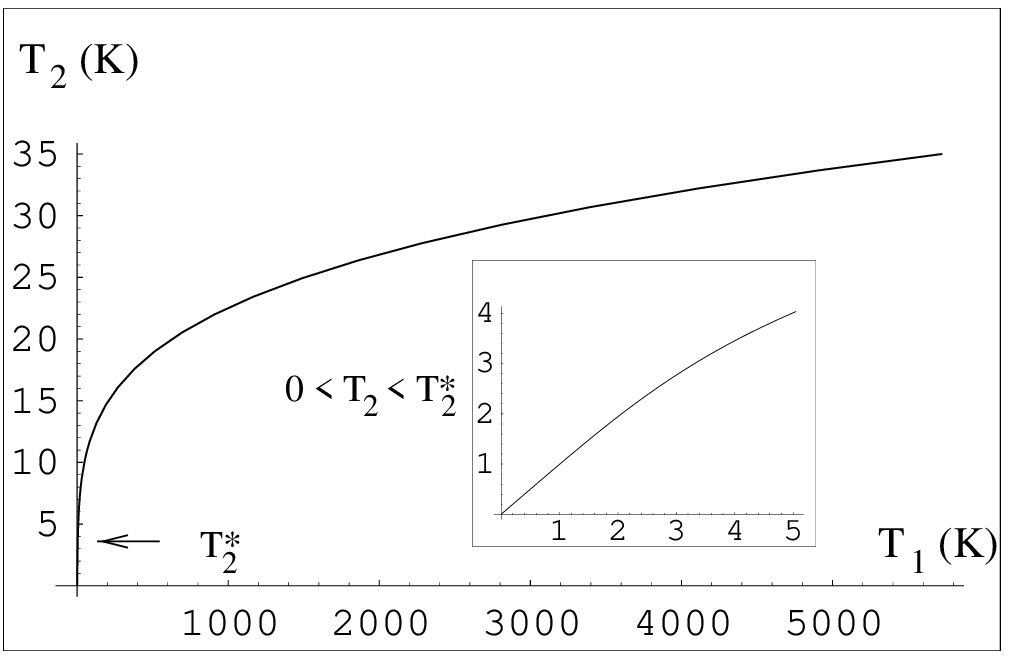}
\caption{ \emph{ {\footnotesize The relationship $T_{1} = T_{1} (T_{2})$ and $T_{2} = T_{2} (T_{1})$.
In this figure we have taken $k$ = 0.3 W m$^{-1}$ K$^{-1}$,
$\Delta R$ = 20 km. It follows from (\ref{eq:iv18}) $T_{2}^{*} \sim 4 \, \text{K}$. 
 \footnotesize }}}  \label{fig:iv1}
\end{center}
\end{figure} \\
Notice that $T_{1}$ and $T_{2}$ have different scales and the linear region $T_{1} = T_{2}$ appears flat.
For $T_{2} < T_{2}^{*}$ we also have that 
$ \nabla_{r} T $
is small. \\
We now proceed in this section adopting an approximation which is not physically restrictive  and simplifies the calculations. We assume that the crust $\Delta R$ of the planet is small with respect to both $R_{1}$ and $R_{2}$
\begin{equation} \label{eq:iv23}
 \frac{R_{2}}{R_{1}} \sim \frac{r}{R_{1}} \sim 1
\end{equation}
Using (\ref{eq:iv23}), equation (\ref{eq:iv11}) becomes linear in $r$:
\begin{equation} \label{eq:iv24}
T  (r,t) = T_{1} (t) - \frac{1}{k} \, \sigma \, T_{2}^{4} (t) \, ( r - R_{1} )
\end{equation}
We have moreover 
\begin{equation} \label{eq:iv25}
\begin{array}{ll}
a \to & a ' = \displaystyle{ \frac{\sigma}{k} \,  \Delta R }, \\
& \\
b \to & b' = \displaystyle{ \frac{c_{v} \, \rho  \, R_{1}}{ 3 \, \sigma} }, \\
\end{array}
\end{equation}
\begin{equation} \label{eq:iv26}
T_{1} = T_{2} + a' \, T_{2}^{4},
\end{equation}
\begin{equation} \label{eq:iv27}
T_{2}^{'*} = \left ( \frac{k}{4 \, \sigma \,  \Delta R}  \right )^{\frac{1}{3}},
\end{equation}
\begin{equation} \label{eq:iv28}
\begin{array}{llr}
\tau_{1} ' = & \displaystyle{ \frac{4}{3} \, \frac{c_{v} \, \rho}{k}  \, \Delta R \, R_{1}}  & \qquad \qquad
\qquad \qquad \qquad \qquad \qquad \qquad a) \\
& \text{and} & \\
\tau_{2} ' = & \displaystyle{ \frac{1}{9} \, (e^{3} - 1) \,
 \frac{c_{v} \, \rho}{\sigma}  \, R_{1} \, \frac{1}{T_{20}^{3}} } & \qquad \qquad b) \\
\end{array}
\end{equation}
From now on we adopt the linear approximation, namely we use the above expressions. 
However we do not continue to carry the primes.

%% file: batt5.tex
\section{Thermal availability in the shell}
\setcounter{equation}{0}
\setcounter{figure}{0}
We have now the ingredients necessary to evaluate the Gibbs availability in the shell.
With reference to \S 3, we use the Carnot availability
\begin{equation} \label{eq:v1}
w = k \, \frac{(\nabla T )^{2}}{T} 
\end{equation}
and we have for our model planet
\begin{equation} \label{eq:v2}
T  (r, t) = T_{1} (t) - \frac{1}{k} \, \sigma \, (r - R_{1} ) \, T_{2}^{4} (t)
\end{equation}
\begin{equation} \label{eq:v3}
\nabla T = \frac{\ud T}{\ud r} = - \frac{1}{k} \, \sigma \, T_{2}^{4} (t)
\end{equation}
so that
\begin{equation} \label{eq:v4}
w (r, t) = \frac{\frac{1}{k} \, \sigma^{2} \, T_{2}^{8} (t)}{T_{1} (t) - \frac{1}{k} \, \sigma \, (r - R_{1} ) \, T_{2}^{4} (t)} 
\end{equation}
We see that $w (r, t)$ increases with $r$ in the interval $R_{1} \le r \le R_{2}$
and  we obtain from (\ref{eq:v4}) this ratio
\begin{equation} \label{eq:v5}
\frac{w (R_{1})}{w (R_{2})} = \frac{T_{2}}{T_{1}}; \qquad \qquad \qquad \text{for any given} \, t
\end{equation}
Let us discuss now the trade-off between availability and thermal mean-life of the host planet. In the foregoing we  consider the regime $T_{2} > T_{2}^{*}$; the mean life $\tau \sim \tau_{2}$ no longer  appears, so we will use the symbol $\tau$
for $\tau_{1}$ in what follows.
   \\
We consider
\begin{equation} \label{eq:v6}
w (r, t) = \frac{\frac{1}{k} \, \sigma^{2} \, T_{2}^{8} (t)}{T (r, t)} 
\end{equation}
in the regime  $T_{2} > T_{2}^{*}$, that means that
\begin{equation} \label{eq:v7}
T_{2} \sim \left ( \frac{T_{1}}{a}  \right )^{\frac{1}{4}},
\end{equation}
so that (\ref{eq:v6}) becomes
\begin{equation} \label{eq:v8}
w (r, t) = \frac{\frac{1}{k} \, \sigma^{2} \, \left ( \frac{T_{1} (t)}{a}  \right )^{2}}{T (r, t)} 
\end{equation}
Using the linear approximation
\begin{equation*} 
 a  = \frac{\sigma}{k} \,  \Delta R  
\end{equation*}
we now have
\begin{equation} \label{eq:v9}
w (r, t) = k  \, \frac{\frac{T_{1}^{2} (t)}{\Delta R^{2}}}{T (r, t)}. 
\end{equation}
At a given $T_{2}$ and for fixed $r$ between $R_{1}$ and $R_{2}$, the dependence of $w$ on $\Delta R$ satisfies the following scaling law.
\begin{equation} \label{eq:v10}
w (\alpha \, \Delta R) = \frac{1}{\alpha^{2}} \, w (\Delta R)
\end{equation}
Always in the regime $T_{2} > T_{2}^{*}$, we have 
\begin{equation} \label{eq:v11}
\tau   =  \frac{4}{3} \, \frac{c_{v} \, \rho}{k}   \, R_{1} \, \Delta R 
\end{equation}
At a fixed $r$ we have
\begin{equation} \label{eq:v12}
\tau (\alpha \, \Delta R) =  \alpha \, \tau (\Delta R). 
\end{equation}
The comparison between (\ref{eq:v10}) and (\ref{eq:v12}) gives the following result: for a given supply of thermal energy stored in the planet, $\frac{4}{3} \, \pi \, c_{v} \, \rho \, R_{1}^{3} \, T_{1} $, a larger availability in the shell implies a thinner shell and therefore a shorter mean-life of the stored energy. 
\\
As far as we know from the properties of terrestrial life, living organisms require a  rather narrow range of ambient temperature.
For example they need liquid water.
The phase diagram of water is known and gives the domain  $T, \, P$ where we have the liquid phase.
To quantify these ideas, we define a best temperature $T_{b}$ and we set
\begin{equation*} 
 T_{b} = 300 \, \text{K} 
\end{equation*}
Fixing $T_{b}$ poses a problem because both $T$ and $w$ are function of $r$ and $t$.
We wish to show that the imposition
\begin{equation} \label{eq:v13}
T (r, t) = T_{b} 
\end{equation}
has implications which can be explicitly calculated.
First of all (\ref{eq:v13}) is an algebraic equation in $r$ which defines a function 
\begin{equation} \label{eq:v14}
r_{b} = r_{b} (t) 
\end{equation}
and, putting (\ref{eq:v13}) into (\ref{eq:v6}), defines also
\begin{equation} \label{eq:v15}
w_{b} = w_{b} (t) 
\end{equation}
In particular we will show that
\begin{equation} \label{eq:v16}
r_{b} (t) < r_{b} (0) \qquad \qquad \qquad \text{for} \quad t > 0 
\end{equation}
\begin{equation} \label{eq:v17}
w_{b} (t) < w_{b} (0) \qquad \quad \quad \text{for} \quad t > 0 
\end{equation}
The inequality (\ref{eq:v16}) implies that the conditions for life must migrate from a location near the surface toward the interior, and correspondingly (\ref{eq:v17}) says that the availability decreases with the migration.
Equation (\ref{eq:v14}) can be written explicitly in the linear approximation
\begin{equation} \label{eq:v18}
r_{b} (t) =  R_{1} +  \frac{(T_{1} (t) - T_{b})}{\frac{\sigma}{k}  \, T_{2}^{4}} 
\end{equation}
We see that (obviously we consider $T_{10} > T_{b}$)
$T_{1}$ reaches the value  $T_{b}$ when $t$ reaches a certain value $\hat{t}$,
namely
\begin{equation*} 
r_{b}  \to R_{1}; \qquad \qquad T_{b}  \to T_{1} \qquad \text{for} \qquad t  \to \hat{t}
\end{equation*}
In other words for $ t  \to \hat{t}$ the location of life reaches the bottom of the crust. From that moment on, $T_{1} (t)$ keeps decreasing and the requirement
$T \simeq T_{b}$ can no longer be maintained. Also the availability can be written explicitly
\begin{equation} \label{eq:v19}
 w_{b} = \frac{\frac{\sigma^{2}}{k} \,  T_{2}^{8} (t)}{T_{b}} 
\end{equation}
and we see at once that $w_{b}$ decreases with $t$. \\
The question is whether $\hat{t}$ is large or not.
We have defined the time $\hat{t}$ with the equation
\begin{equation} \label{eq:v26}
T_{1} (\hat{t}) = T_{b} 
\end{equation}
Using
\begin{equation} \label{eq:v27}
T_{1} (t) \sim a \, T_{2}^{4} (t) = a \, T_{20}^{4}   \, e^{- 4 \, \frac{t}{\tau}}  
\end{equation}
we obtain
\begin{equation} \label{eq:v28}
\hat{t} \sim \frac{\tau}{4} \, \ln \left ( \frac{a \, T_{20}^{4}}{T_{b}} \right ) = 
\frac{\tau}{4} \, \ln \left ( \frac{T_{10}}{T_{b}} \right )  
\end{equation}
Notice that
\begin{equation} \label{eq:v29}
\hat{t} = 0 \qquad \qquad \text{if} \qquad \qquad T_{10} = T_{b}
\end{equation}
\begin{equation} \label{eq:v30}
\hat{t} \quad { \small { > \atop < }} \quad \tau  \qquad \qquad \text{if} \qquad \qquad \left \{
\begin{array}{l}
\displaystyle{ \ln \left ( \frac{T_{10}}{T_{b}} \right ) > 4 } \\
\\
\displaystyle{ \ln \left ( \frac{T_{10}}{T_{b}}  \right ) < 4 } \\
\end{array} \right .
\end{equation}
We can write the time dependence of $r_{b}$ and $w_{b}$ explicitly, showing how it is  affected by the model parameters. \\
We are in the $T_{2} > T_{2}^{*}$ regime, so that
\begin{equation} \label{eq:v20}
T_{2} (t) \sim T_{20}  \, e^{- \frac{t}{\tau}}  
\end{equation}
with 
\begin{equation} \label{eq:v21}
\tau \sim  \frac{4}{3} \,   \frac{c_{v} \, \rho}{k}  \, \Delta R \, R_{1}
\end{equation}
Using (\ref{eq:v20}) in (\ref{eq:v18}) and (\ref{eq:v19}) we get (for $0 < t < \hat{t}$):
\begin{equation} \label{eq:v22}
\begin{split}
r_{b} (t)  & = R_{1} +  \frac{k \, a}{\sigma} - \frac{k}{\sigma} \,\frac{T_{b}}{T_{20}^{4}} \, e^{4 \, \frac{t}{\tau}} \\
& = R_{1} +  \frac{k \, a}{\sigma} \, \left ( 1 - \frac{T_{b}}{T_{10}} \, e^{4 \, \frac{t}{\tau}} \right ) \\
\end{split}
\end{equation}
and
\begin{equation} \label{eq:v23}
w_{b} (t)   = \frac{\sigma^{2}}{k} \, \frac{T_{20}^{8}}{T_{b}} \, e^{- 8 \, \frac{t}{\tau}} 
\end{equation}
It turns out, by inspection of (\ref{eq:v22}) and (\ref{eq:v23}) that the crucial parameter is $\tau$.
 Knowing the value of $\tau$, we can estimate the migration length
$r_{b} (0) - r_{b} (t) = \delta r (t)$ and the availability decrease $w_{b} (0) - w_{b} (t) = \delta w (t)$ 
after a certain time $t$. From (\ref{eq:v22}) we have
\begin{equation} \label{eq:v24}
\delta r_{b} (t)  =  \frac{k}{\sigma} \,\frac{T_{b}}{T_{20}^{4}} \, \left ( e^{4 \, \frac{t}{\tau}} - 1 \right ) 
\end{equation}
while from (\ref{eq:v23}) we have
\begin{equation} \label{eq:v25}
\delta w_{b} (t)  =   \frac{\sigma^{2}}{k} \, \frac{T_{20}^{8}}{T_{b}} \, \left ( 1 - e^{- 8 \, \frac{t}{\tau}} \right ).
\end{equation}
Note that the linearization of $T (r, t)$ and the limitation to $T_{2} > T_{2}^{*}$ are reasonable simplifications.
Obviously $r_{b} (t)$ and $w_{b} (t)$ could be calculated numerically choosing the parameters of the model and the initial condition
$T_{20}$ at will, and without the above approximations. \\ \quad \\
We conclude this section by giving a numerical estimate of $r_{b}$,  $w_{b}$, $\delta r_{b}$,  $\delta w_{b}$ 
and $\hat{t}$ for a certain set of the planetary parameters. We take a planet that vaguely resembles the Earth.
\quad \\ \quad \\
\emph{Material parameters} \\ 
\begin{tabular}{lll} \label{tab:uno}
$c_{v}$  & $5 \cdot 10 ^{3}$ J kg$^{-1}$ K$^{-1}$ & Iron \\ 
$\rho$   & $10 ^{4}$  kg m$^{-3}$ & Iron \\
$k$  &0.3 W m$^{-1}$ K$^{-1}$ & Sand soil \\
\end{tabular} \quad \\ \quad \\
\emph{Geometry} \\ 
\begin{tabular}{ll} \label{tab:due}
$R_{1}  $  & 6.28 $\cdot 10^{6}$ m \\
$R_{2}  $  & 6.3 $\cdot 10^{6}$ m \\
$\Delta R = R_{2} - R_{1} $  & 20 km \\
\end{tabular} \\ \quad \\
$\sigma$  is the Stefan-Boltzmann constant $\sim 5.67 \cdot 10^{-8}$ W m$^{-2}$ K$^{-4}$, 
$T_{b} =  300$ K by our choice. \\
\emph{Initial condition} \cite{ref:peo}:
\begin{equation*} 
q_{0}   \sim 0.05 \frac{\text{W}}{\text{m}^{2}}  
\end{equation*}
\emph{Note: The Earth is about $10^{9}$ years old. Here we take the present age as the initial time. Alternatively we could take a higher $q_{0}$ and correspondingly higher $T_{10}$ and begin our clock much earlier. The reader can play with these numbers.} \\
Since
\begin{equation*} 
q_{0} = \sigma \,T_{20}^{4}; \qquad \qquad  \text{we have}  \qquad   \qquad 
T_{20} = \left (\frac{q_{0}}{\sigma} \right )^{\frac{1}{4}} \sim 30.6 \, \text{K}; 
\end{equation*}
and
\begin{equation*} 
 T_{10} \sim a \,
T_{20}^{4} = 3314 \, \text{K}
\end{equation*}
\begin{equation*} 
 a = \frac{\sigma}{k} \, \Delta R = 3.78 \cdot 10^{-3} \, \text{K}^{-3}
\end{equation*}
We calculate  $r_{b} - R_{1}$ for $t = 0$
\begin{equation} \label{eq:v32}
r_{b} (0) -  R_{1} =  \frac{k}{\sigma} \, a \, \left ( 1 -  \frac{T_{b}}{T_{10}} \right ) 
 \sim 18.2 \cdot 10^{3} \, \text{m}
\end{equation}
which means $r_{b}$ is $\sim 2$ km  below the surface at $r = R_{2}$, and  $w_{b}$ for $t = 0$ is
\begin{equation} \label{eq:v33}
 w_{b} = \frac{\frac{\sigma^{2}}{k} \,  T_{20}^{8}}{T_{b}} \sim 2.75 \cdot 10^{- 5} \, \frac{\text{W}}{\text{m}^{3}}.
\end{equation}
Concerning the time dependence we calculate $\tau$ and get
\begin{equation} \label{eq:v34}
 \tau  \sim  \frac{4}{3} \, \frac{c_{v} \, \rho}{k}   \, R_{1} \, \Delta R  \sim 8.85 \cdot 10^{11} \, \text{yr}.
\end{equation}
Given $\tau$, using equation (\ref{eq:v28}) we calculate $\hat{t}$
\begin{equation} \label{eq:v35}
\hat{t} = 
\frac{\tau}{4} \, \ln \left ( \frac{T_{10}}{T_{b}} \right ) \sim 5.3 \cdot 10^{11} \, \text{yr}  \sim 0.6 \, \tau.
\end{equation}
What are the quantities $\delta r_{b}$ and $\delta w_{b}$ after $t = 10^{6}$ yr? We get from (\ref{eq:v24}) and (\ref{eq:v25})
\begin{equation} \label{eq:v36}
\delta r_{b} (10^{6} \, \text{yr} )  = 
 \frac{k}{\sigma} \,\frac{T_{b}}{T_{20}^{4}} \, \left ( e^{4 \, \frac{t}{\tau}} - 1 \right ) \sim 0.008 \, \text{m}
\end{equation}
From (\ref{eq:v23}) we have
\begin{equation} \label{eq:v37}
\delta w_{b} (10^{6} \, \text{yr}  )  =  
 \frac{\sigma^{2}}{k} \, \frac{T_{20}^{8}}{T_{b}} \, \left ( 1 - e^{- 8 \, \frac{t}{\tau}} \right ) \sim 2.5 \cdot 10^{-10}
\, \frac{\text{W}}{\text{m}^{3}}
\end{equation}
The faintness of the continuous availability evaluated in the above numerical example is an indication of a general property. In fact we cannot make wild hypotheses with regards to $T_{20}$, $\Delta R$ and the core equation of state. In particular, $\Delta R$ governs the trade off between the planet's thermal life, $\tau$, and the availability $w$ that it can house.
It is therefore obvious that $w$ is in general small. \\
We may consider the possibility of surface availability, rather than volume availability. This implies a discontinuity of $T$, and $\nabla T$, and this may exist in the presence of a small scale structure on the shell of the planet, for instance some form of hot streams coming from the core and protunding into the cooler crust. In this case we may have an interface between the stream at $T_{1}$
and the ambient at $T_{2}$. Such a discontinuity produces an availability
\begin{equation} \label{eq:v38}
w = \vec{q} \cdot \vec{\xi} \, \left ( \frac{T_{1} - T_{2}}{T_{1}} \right ) \, \delta (\xi - \xi_{0})
\end{equation}
where we make reference to the following figure: 
\begin{center}
\includegraphics[width=6 cm]{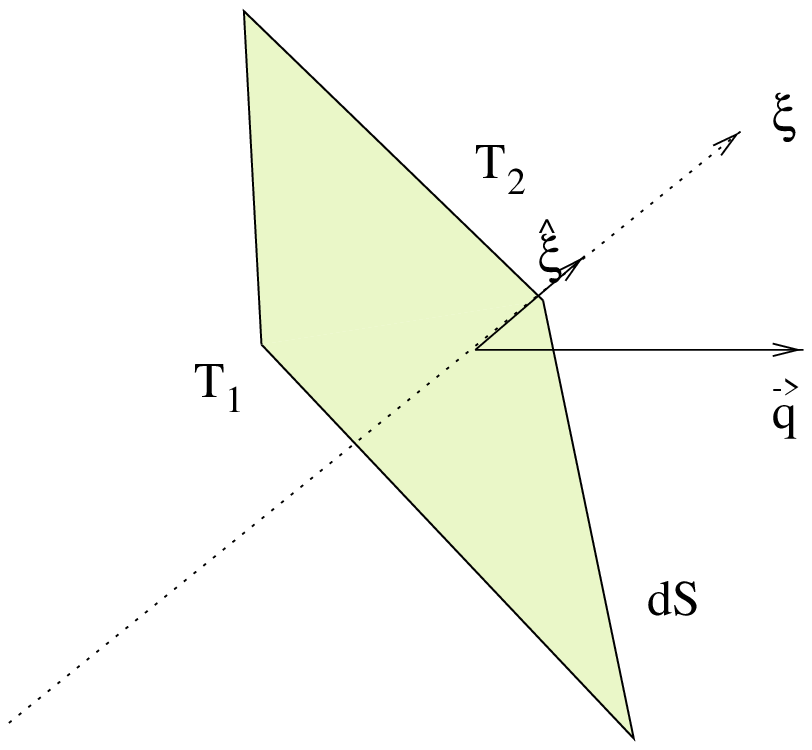} 
\end{center}
The quantity $\vec{q}$ is the heat flux, $\hat{\xi}$ is the normal vector, the straight line $\xi$
being perpendicular to the surface element $\ud S$,
$T_{1}$ is the stream temperature, $T_{2}$ the ambient temperature. \\
The justification for considering the surface availability  (\ref{eq:v38}) is that temperature discontinuities exist on Earth.
We have various examples of hot springs, both at the interface solid-atmosphere and solid-ocean. We may reasonably think that this kind of structure may exist also on other planets. \\
On Earth, with such discontinuities there sometimes corresponds  a local community of living organisms, an opportunistic niche. 
Can we make the hypothesis that such niches exist on planets without a  star? This is not an innocent hypothesis, because it implies the preliminary verification of a very strong statement, namely that the existence of the terrestrial niches attached to discontinuities
is disjoint from the existence of the global living community (which for our planet is a photon driven system)
both in an ecological and evolutionary sense.
In the absence of a proof of this statement, the consideration of life niches belonging to discontinuities on planets without star is not a tenable hypothesis, but a belief.

%% file: batt6.tex
\section{Planet with star. The global steady state}
\setcounter{equation}{0}
\setcounter{figure}{0}
A planet that receives radiation from a star has on its surface a thermal distribution much different from that of a planet without star.
What we have discussed in \S 4 is a global transient with spherical symmetry.
The interesting dynamics is in the insulating shell.  \\
If a planet is sufficiently close to its star, the radiation flow 
 impinging on the planet can be stronger than the energy
 flow originated by the planet's core, and the planet's  surface dynamics becomes dominated by the external input. \\
\emph{Note. Numbers for the Earth: power flow from the core
$q_{0} = 0.05 \, \frac{\text{W}}{\text{m}^{2}}$,
solar constant $q^{\gamma} = 1350 \, \frac{\text{W}}{\text{m}^{2}}$}. \\
The external  input is no longer symmetric in the reference system of the planet. \\
On the other hand, if the total flow investing the planet is constant, the surface of the planet can be in a global steady state. 
The first question is therefore with regards to the astronomical conditions necessary to have a global steady state. \\
In the solar system the planetary orbits are in good approximation elliptic and also with small eccentricity. A good astronomical question is whether this architecture is unique to the Sun and its planets or is a general Newtonian property. We leave the deep and difficult problem of  the stability and structure of planetary systems to the astronomers.
We remark, however, that in the absence of near circular orbits the radiation flux hitting a planet may be a chaotic  (non periodic) function of the time \cite{ref:alf1}, and the thermal distribution, created on the surface of the planet, will be something which is neither a transient nor a steady state.  \\
Consider now a near circular orbit. The next question concerns the temperature 
of the star $T_{s}$, with radius $R_{s}$ and the distance star-planet $d$. In fact from these three numbers follows the average surface temperature of a  planet with a black body surface and without an atmosphere. The average temperature on the circular orbit of radius $d$ is
calculated from the energy balance. In this formula $\phi^{in}$ is the portion of the isotropic total radiation emitted by the star and intercepted by the cross section of the planet $\pi \, R^{2}$.
$\phi^{out}$ is the total energy radiated by the planet. 
\begin{equation} \label{eq:vi1} 
\phi^{in} = \pi \, R^{2} \, \sigma \, T_{s}^{4} \, \frac{R_{s}^{2}}{d^{2}} =
\phi^{out} = 4 \, \pi \, R^{2} \, \sigma \, T_{av}^{4}
\end{equation}
From the balance (\ref{eq:vi1}) we get $T_{av}$:
\begin{equation} \label{eq:vi2}
T_{av} = \frac{1}{\sqrt{2}} \, \sqrt{\frac{R_{s}}{d}} \, T_{s}
\end{equation}
Now $T_{av} \ll T_{s}$ because the radiation flow originated on the star is diluted at the distance $d$; but the spectroscopic temperature remains $T_{s}$ \cite{ref:nc}. \\
In turn, if we look for long stellar lifetimes inside the main sequence, $T_{s}$, $R_{s}$ and
$\tau_{s}$ are related to the stellar mass. If we require a long duration of stability, for instance $\tau_{s} \sim
10^{9} \, \text{yr}$, the values of $T_{s}$ and $R_{s}$ are not wildly different from those of the Sun,
so the consideration of $T_{s}$ in the order of few thousand K, makes sense. \\
A global steady state takes place if
the global balance (\ref{eq:vi1}) is satisfied with a surface temperature $T (\theta, \varphi, t)$ which depends on position and time.
The detailed properties of the global steady state are related to the total amount of greenhouse gases.
 In fact $\phi^{out}$ can differ from $\phi^{in}$
during a transient. This transient can only be a period of increase or decrease of greenhouse
molecules in the planetary atmosphere.
Now, except for external causes, primarily planetary collisions, there is no good reason to believe that the total amount of greenhouse gas must oscillate. It follows that a global non equilibrium steady state is the rule.
If the amount of greenhouse gas is constant, the effective average surface temperature $T_{av}^{eff}$ will be different from the average temperature $T_{av}$, calculated above, that takes into account only the astronomical parameters.
We can in general say  that
\begin{equation*}
T_{av}^{eff} = \gamma \, T_{av} 
\end{equation*}
with $\gamma > 1$. The value of $\gamma$ depends on the history of the planet's formation, and the actual amount of greenhouse gases existing at a certain stage of its evolution.

%% file: batt7.tex
\section{Photon availability}
\setcounter{equation}{0}
\setcounter{figure}{0}
At present we do not have any ideas about the possibility of gaseous life. Life is structured, shaped; gases are, by definition, interactionless
and shapeless.
So we disregard  it here and  consider the photons that have passed through the atmosphere and hit a liquid or solid
layer of molecules that can thermalize. We may assume that only a few collisions are sufficient to thermalize the radiation flow, or, in other words, the thermalization is a surface effect.
The surface temperature changes with  time and the latitude on the planet: day and night, and the seasons. This is a very important fact for the understanding of the complexity of photon-powered life.
$T (\vec{r},t)$ oscillates around the value $T_{av}^{eff}$,  the average value that takes into account the greenhouse effect. \\
The photon availability $w^{\gamma}$ is rather different from the conduction availability, $w^{carnot}$, studied in the preceding sections.
In fact, considering that the layer involved in the thermalization of the photons is very thin, we may write
\begin{equation} \label{eq:vii1}
w^{\gamma} (r) = q^{\gamma} \, \eta \, \delta (r - R)
\end{equation}
($\delta$ is the Dirac function, the symbol $\gamma$ stands for photon). \\ 
$[w] = \frac{\text{W}}{\text{m}^{3}}$, \qquad  $[\delta (r)] = \text{m}^{-1}$. \\
In (\ref{eq:vii1})
\begin{equation} \label{eq:vii2}
q^{\gamma} = \frac{R_{s}^{2}}{d^{2}} \, \sigma \, T_{s}^{4} \cdot f (\theta, \varphi, t)
\end{equation}
where $f (\theta, \varphi, t)$ is a kinematic factor depending on latitude and longitude.
The factor $f (\theta, \varphi, t)$ takes into account the relative motion between star and planet.
See ref. \cite{ref:nc} for details.
The efficiency $\frac{\Delta T}{T}$ is given by
\begin{equation} \label{eq:vii3}
\eta = \frac{\Delta T}{T} = \frac{T_{s} - T (R)}{T_{s}}
\end{equation}
The efficiency $\eta$ is an astronomical property rather than a planetary one. In fact, using for $T (R)$ the value $T_{av}$
(\ref{eq:vi2})
\begin{equation} \label{eq:vii4}
T (R) = T_{av} = \frac{1}{\sqrt{2}} \, \sqrt{\frac{R_{s}}{d}} \, T_{s}
\end{equation}
we get
\begin{equation} \label{eq:vii5}
\eta = 1 - \frac{1}{\sqrt{2}} \, \sqrt{\frac{R_{s}}{d}}
\end{equation}
We see immediately that
\begin{equation} \label{eq:vii6}
w^{\gamma} (r) = \frac{R_{s}^{2}}{d^{2}}  \cdot f (\theta, \varphi, t) 
\, \sigma \, T_{s}^{4} \, \left (  1 - \frac{1}{\sqrt{2}} \, \sqrt{\frac{R_{s}}{d}}   \right )
\, \delta (r - R)
\end{equation}
Equation (\ref{eq:vii6}) says that the photon availability is an astronomical property, and the planet radius appears uniquely in the delta function $\delta (r - R )$.
The photon availability has several very important properties.
\begin{description}
\item[-] $T_{s}$ is high, and is in a  range that may involve  important chemical reactions
\item[-] The power input $w^{\gamma}$ works mainly on the surface,
\item[-] There is in general a high value of $\eta$, namely $\eta \sim 1$, or, in other words $T_{s}$ and $T_{av}$ are quite different from each other.
\end{description}
As citizens of the planet Earth, we are the best observers of life powered by photon availability. We see that
\begin{description}
\item[-] An organism powered by photons is richly structured. This is a direct consequence of the fact that the living organism must develop itself behind the receiving surface: it must be organized in a structure.
\item[-] The photon-driven chemical reactions constituting the input to the complex state $C$, can be, in principle, the 
alternative to mining
of chemical components. This alternative  is indeed realized by Nature. In fact
 it happens that the complex living state $C$, which does not need mining in the input, does not need a reservoir for the outgoing chemical states either.
\end{description}
We have described the metabolism based on mining with the graph (\S 2)
\begin{center}
\includegraphics[width=8 cm]{bat1.eps}   
\end{center}
The complex state $C$ of the terrestrial ecosystem belongs to a different graph:
\begin{center}
\includegraphics[width=4.5 cm]{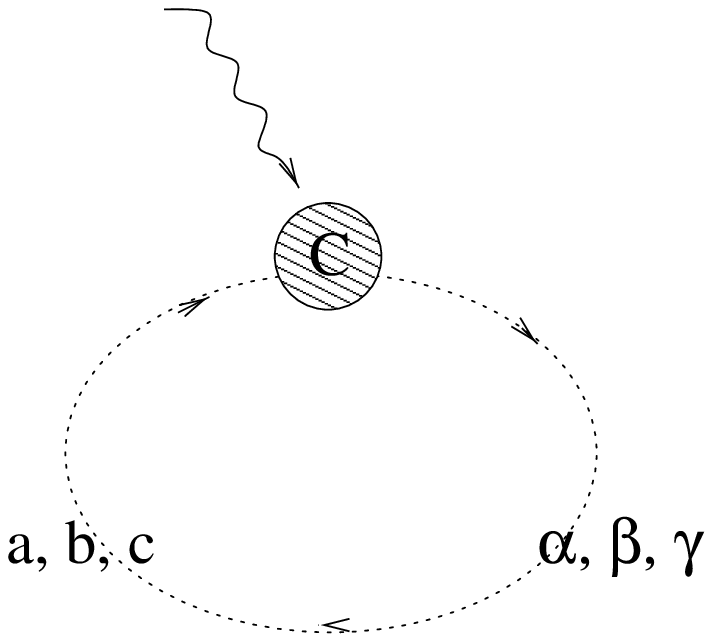}   
\end{center}
On the surface of the planet Earth the outgoing components $\alpha, \, \beta, \, \gamma$ circulate in the physical ambient and perform a cycle reproducing exactly $a, \, b, \, c$ in the amount that is required by $C$. It is uniquely by closing the cycle
$a, \, b, \, c \to \alpha, \, \beta, \, \gamma $ that life on a planet can be eternal, where eternal means over the lifespan of the
host star. \\
Notice that the cycle has an upper limit to its total power and this is the integral of the photon availability on the planet's surface
\begin{equation} \label{eq:vii7}
\Phi^{max} \simeq \int_{S} q (\theta, \varphi, t) \, \frac{T_{s} - T (\theta, \varphi, t)}{T_{s}}
  \, \ud S; \qquad \qquad [\Phi^{max}] = \text{W}
\end{equation}
 $T (\theta, \varphi, t)$ is the temperature field on the planet's surface.
Obviously the total availability $\Phi^{max}$   is utilized by the population of the living organism only in part.
The other part goes into the physical channel, mainly the non-organic water cycle and  thermal circulation.
We know that  on Earth about $10^{-2}$ of the total incoming flux goes into the life channel \cite{ref:mar}.
This ratio comes from observation, not from theory. But the important fact is that life based on stellar
 radiation and organized according to the above cycle has a physical upper limit to its growth, and consequently its duration \emph{is not population dependent}, as it happens with life from mining
(\S 2) but \emph{star dependent}.

%% file: batt8.tex
\section{Conclusion}
Comparing the numerical discussion of the thermal availability in \S 5 and the photon availability in \S 7, it appears
that the physical chances for life on a planet without a star are rather miserable in comparison to a planet in a near circular orbit around a star. \\
In both cases, the time scales for the endurance of a hot planetary core, or the burning of the star, are sufficiently long; that is not the limiting effect for the permanence of life. It is the material cycle (\S 1) that matters. It is conceivable
that when the available energy is very low, the strategy of life goes into the direction of mining, for both chemical elements and energy.
If this is the choice, life becomes a transient. It may exist by bursts,
 and chances of communication with other bursts in other places in the Universe may become negligible. Life,
as developed on Earth functions according to a global cycle, and therefore the duration of life on the planet
can be equal to the duration of its host star (\S 7).
\\
The understanding of the life-ambient cycle is a much harder problem than the mere evaluation of the availability. It is reasonable
to think that the existence of this cycle, and the corresponding permanence of life, may pose other very severe  conditions
for the possibility of life in the Universe.
\\
At present we are not able to formulate these conditions: we are only able to witness that they exist on Earth.
\section*{Aknowledgements}
We wish to thank Prof. Guido Cosenza for several stimulating discussions, and Dr. Ronald Drimmel
for critical reading of the manuscript.